\documentclass[9pt, a4paper]{article}
\usepackage{geometry}
\usepackage{graphicx}
\usepackage{array}
\usepackage{times}
\usepackage{enumitem}
\usepackage{algorithm}      
\usepackage{amsmath}        
\usepackage{amssymb}        
\usepackage{makecell}
\usepackage{algpseudocode}
\usepackage{booktabs}
\usepackage{multirow} 
\usepackage{subcaption} 
\usepackage[colorlinks=true, linkcolor=blue, citecolor=blue, urlcolor=red]{hyperref}
\usepackage[numbers, sort&compress]{natbib} 
\newtheorem{definition}{Definition}
\newtheorem{lemma}{Lemma} 
\newtheorem{theorem}{Theorem}
\title{\textbf{\large An Improved Quasi-Physical Dynamic  Algorithm for Efficient Circular Coverage in Arbitrary Convex Polygons}}
\author{
	Zeping Yi\textsuperscript{1},
	Yongjun Wang\textsuperscript{$1^*$},
	Baoshan Wang\textsuperscript{1},
	Songyi Liu\textsuperscript{1}
	\\
	\small \textsuperscript{1}School of Mathematical Sciences, Beihang University, Beijing, 100191, China
	\\
	\small *Corresponding author(s): wangyj@buaa.edu.cn;\\
	\small Contributing authors: yzping@buaa.edu.cn; bwang@buaa.edu.cn;liusongyi@buaa.edu.cn
}

\date{}

\begin{document}
	\maketitle
	\thispagestyle{empty}  
	\begin{abstract}
	The optimal circle coverage problem aims to find a configuration of circles that maximizes the covered area within a given region. Although theoretical optimal solutions exist for simple cases, the problem's NP-hard characteristic makes the problem computationally intractable for complex polygons with numerous circles. Prevailing methods are largely confined to regular domains, while the few algorithms designed for irregular polygons suffer from poor initialization, unmanaged boundary effects, and 
	excessive overlap among circles, resulting in low coverage efficiency. Consequently, we propose an Improved Quasi-Physical Dynamic(IQPD) algorithm for arbitrary convex polygons. Our core contributions are threefold: (1)proposing a structure-preserving initialization strategy that maps a hexagonal close-packing of circles into the target polygon via scaling and affine transformation; (2)constructing a virtual force field incorporating friction and a radius-expansion optimization iteration model; (3)designing a boundary-surrounding strategy based on normal and tangential gradients to retrieve overflowing circles. Experimental results demonstrate that our algorithm significantly outperforms four state-of-the-art methods on seven metrics across a variety of convex polygons. This work could provide a more efficient solution for operational optimization or resource allocation in practical applications.\\
		
	\end{abstract}

\newpage

\clearpage  
\pagenumbering{arabic}  
\setcounter{page}{1}    
\section{Introduction} 

 Given a set of circles and a polygonal region, how to arrange them to maximize the coverage area of the given region or even achieve full coverage? Although this problem seems simple, it has a wide range of applications in our life. In practical scenarios such as emergency rescue, cutting manufacturing \cite{HIFI2004675}, agricultural irrigation\cite{CASTILLO2008786}, and facility location selection \cite{Kumar2024} , determining the optimal arrangement of circles to achieve a larger coverage area is of great practical value.
 
 This problem originates from a famous close packing problem \cite{1939,2017}, which seeks to place the maximum number of circles with known radii inside a given polygon without overlap. As a classic problem in computational geometry and operations research optimization, it is typically NP-hard \cite{2023,20241,2021},and provably optimal solutions are known only for specific cases (certain simple shapes, specific quantities, or specific radii). Consequently, heuristic algorithm becomes a predominant approach for obtaining near-optimal arrangements \cite{2025,2016}. As early as the time of Archimedes, scholars began investigating optimal arrangements of circles in the plane, though the close packing problem had not yet developed into a formal mathematical theory\cite{Netz2005}. In the 17th century, Johannes Kepler discussed the close packing of congruent circles on a plane in his work $On\ the\ Six-Cornered\ Snowflake$, and conjectured that hexagonal packing of spheres would be the densest configuration in three-dimensional space\cite{2011}, but he did not provide a proof for the optimality of two-dimensional circle tiling. Axel Thue made the first attempt to prove that the hexagonal arrangement is the densest tiling method for circles on a plane in 1890 \cite{thue1892om}, but the proof was still incomplete even after being revised in 1910 \cite{thue1910om}. Finally, it was not until 1949 that László Fejes Tóth gave a rigorous proof \cite{toth1949dichteste} that the hexagonal arrangement is the optimal solution for circles packing on a plane (with a density of \(\frac{\pi}{2\sqrt{3}} \approx 0.9069\)). This conclusion has provided valuable experience for researchers to study the optimal coverage problem of irregular polygons \cite{2016,2021,RYU2020125076}.
 
Research on the optimal arrangement of circles mainly falls into two canonical categories: the covering and the packing problem \cite{RYU2020125076, Szabo2007NewApproaches}. The covering problem typically aims to cover a given region completely using a set of congruent circles. The objective is often to minimize the radius or number of circles used for covering or maximize the area of region covered totally, and overlaps between circles are generally permitted. In contrast, the packing problem involves placing a set of circles(the radii are not necessarily equal) within a given container without any overlap. The goal is usually twofold: either to maximize the number or the total area of packed circles inside a fixed container, or to minimize the size of the container (e.g., the radius of a circular container or the area of a rectangle) required to accommodate all circles.
 
Solutions for circle covering and packing problems  can be broadly categorized into two paradigms\cite{RYU2020125076}.The first is the analytic approach, which employs mathematical theory to derive provably optimal arrangements\cite{toth2005thinnest,MELISSEN2000149}. However, due to its computational complexity, this approach is generally feasible only for a small number of circles or highly regular configurations.Therefore,the second paradigms, which is more widely applicable and relies on heuristic algorithms, can be categorized by a diverse set of strategies:1)Quasi-physical simulation\cite{WU2002341,Huang2011},which model circles as repulsive objects, adjusting their positions through collision detection and elastic movement until a stable, dense state is reached.2)Stochastic search techniques,which explore the solution space through guided randomization,such as  genetic algorithm \cite{2016, WANG2018196} and simulated annealing algorithm\cite{2004A, TOLE2023109004}.3)Nonlinear programming method,which leverage geometric structures(e.g.,Voronoi diagram) \cite{RYU2020125076} to frame the problem as an optimization model and is solved with the aforementioned heuristic algorithms eventually.

Nevertheless, in practical circular layout problems, such as agricultural irrigation and wireless communications, the complexity often extends beyond the classical paradigms of pure packing or covering. A more prevalent scenario presents a significant challenge: the number and size of circles are often insufficient for non-overlapping packing yet also inadequate for complete coverage. This situation defines the optimal coverage problem for circles. The core of this problem is to find a near-optimal layout that maximizes the coverage ratio of the target area under the condition that overlaps are permitted, thereby achieving the optimal coverage efficiency with limited resources.

To formulate this problem precisely, we establish the following mathematical model. Given a polygonal region $P$, a circle radius $r$, a number of congruent circles $n$, and center coordinates \((x_i, y_i)\ (i = 1, \ldots, n)\), the covered area is defined as 
$$A_{cover} = Area\left( P \cap \left( \bigcup_{i=1}^n C_i \right) \right),$$ 
where \(C_i\) denotes the circle with center \((x_i, y_i)\) and radius $r$. The mathematical formulation of the circle optimal covering problem is as follows:
\begin{equation}
	\underset{\left( x_1,y_1 \right) ,\cdots ,\left( x_n,y_n \right)}{maximize}A_{cover}/Area(P)
\end{equation}
\begin{equation}
	s.t.
	\quad \left\{ \begin{array}{l}
		\left( x_i,y_i \right) \in P\quad \forall i,\quad dist\left( \left( x_i,y_i \right) ,\partial P \right) \geq 0\\
		C_i\cap P\ne \oslash\\
	\end{array} \right. .
\end{equation}
Here, \( dist \left( (x_i, y_i), \partial P \right) \) denotes the signed distance from the center \( (x_i, y_i) \) to the boundary \( \partial P \) of the polygon. The constraint \( \text{dist}\left( (x_i, y_i), \partial P \right) \geq 0 \) ensures that all circle centers lie within the polygon \( P \). The second constraint, \( C_i \cap P \neq \emptyset \), guarantees that each circle contributes to the coverage of the region.

The objective function \( A_{cover}/Area(P) \) is nonconvex and nonlinear, with numerous local optima.  
The search space defined by all circle centers $(x_i,y_i)$ expands rapidly with \( n \),  
compounding the problem's computational difficulty. Fortunately, the optimal circle coverage problem shares characteristics with both packing and covering problems, allowing many existing methods and strategies from these domains to be adapted. Due to the problem's complexity, globally optimal solutions are generally attainable only for small \( n \) or  
simple polygons \( P \). For larger \( n \) or more complex regions, metaheuristic algorithms are employed to approximate the optimum.

Many metaheuristic algorithms\cite{CHOWDHURY2021102660,biomimetics10110750,Yang2025,Shaikh2025,S01102024,LI2025111431} can achieve high coverage rates in regular polygons, but they often encounter issues such as excessive overlap between circles, a number of circles exceeding the boundary, low coverage rates and long run time when dealing with complex polygons. Moreover, due to the presence of concave angles in non-convex polygons, several complexities arise: non-convex constraints must be handled during the feasibility check of circle centers, the optimization of circle layout must consider whether circles should be embedded into concave cavities, and the calculation of the covered area must address more intricate boundaries and multiple non-connected regions. These factors significantly increase the complexity of the problem. To focus on the study of core coverage mechanisms and establish a stable and effective solution framework, this paper limits its scope to the optimal coverage of circles in arbitrarily complex convex polygons. However, a noticeable gap remains: the well-established quasi-physical simulation framework, proven effective in circle packing, has not been systematically adapted or advanced for the optimal covering problem. Key aspects such as intelligent initialization strategies and iterative refinement within the quasi-physical simulation remain unexplored. While the current research strongly favors applying novel metaheuristics to the optimal circle covering problem, this focus may overlook the potential advantage of well-established strategies or methods for classical covering and packing problems. Through adequate improvement and refinement, many of  them could obtain better solutions than metaheuristics for the optimal circle covering problem. Therefore, we aim to develop an improved quasi-physical algorithm with new coverage strategies applicable to general convex scenarios.
 
This paper is structured as follows: Section 2 
review related work with this paper. Section 3 introduces fundamental concepts in computational geometry. Section 4 elaborates on the improved quasi-physical dynamic algorithm. 
Section 6 concludes the paper and suggests future work.

\section{Literature Review}
 \subsection{Analytic Approach}
 
 Analytic approach is often employed to solve covering problems. Researchers initially focused on the optimal arrangement of $n$ unit circles to cover a circular region of the maximum possible radius \cite{1915}. Subsequent research established optimal arrangements for specific numbers of circles: Bezdek for $n=5,6$ \cite{20241} from 1979 to 1984, and Fejes Tóth for $n=8,9,10$ \cite{toth2005thinnest} in 2005.
 
 Meanwhile, researchers also investigated the coverage of rectangular regions using unit circles. Optimal coverings were progressively determined for $n \leq 5$ \cite{Heppes1997} in 1997, $n \leq 7$ \cite{MELISSEN2000149} in 2000, and eventually $n \leq 12$ \cite{20241} in 2014. To address more general shapes, Gul et al.\cite{2021} studied thin convex domains in 2021, proposing efficient strategies based on single-row and irregular hexagonal grid coverage, complete with theoretical bounds of required circle numbers. Recently, Fekete et al.\cite{20241} provided a closed-form solution for the critical covering area of $\lambda \times 1$ rectangles in 2024, offering a tight theoretical benchmark.
 
 Research also extended to triangular regions. In 1997, Melissen provided optimal and conjectured coverings for equilateral triangles with $n \leq 18$ unit circles\cite{Melissen1997Loosest}. As $n$ increased, determining provably optimal configurations became computationally prohibitive, leading Nurmela et al.\cite{Nurmela01012000} to propose and refine conjectures for up to $n=36$ circles in 2000.
 
 Although substantial progress has been made in covering regular regions, practical applications typically involve irregular polygonal domains. In 2023, Choi et al.\cite{2023} developed an \(O(n \log n)\)-time algorithm for covering any convex polygon with two congruent circles of minimal radius, surpassing prior methods \cite{19981, Kim2000299}. More broadly, Birgin et al.\cite{1012872024} investigated the problem of covering a two-dimensional bounded region using a fixed number of congruent circles with the minimum radius. For a specific class of non-smooth domains, they derived bounds on the optimal radius and provided an asymptotic expansion of the bounds on the optimal  radius as the number of circles tends to infinity.  
 
 The work of Choi et al., bridging theoretical analysis and algorithmic design, represents a significant advance in computational efficiency. However, as problem complexity increases, whether through more covering circles, higher dimensions, or non-convex regions, proving optimality theoretically becomes increasingly challenging. Consequently, researchers turned to adopting heuristic algorithms to solve these more complex covering and packing problems.
 
 \subsection{Heuristic Algorithm with Special Rules}
 
 Heuristic algorithms are primarily employed to address the large-scale equal circle packing problem initially proposed by Kravitz in 1967\cite{Kravitz01031967}. This problem involves packing $n$ congruent circles with radii $r$  into the smallest circular container without overlap, where $R_0$ denotes the radius of the smallest circular container, and $(x_i, y_i)$ represents the coordinates of the center of the $i$-th circle. To simplify the model, we can assume that the circular container is centered at the origin $(0, 0)$. Therefore,the mathematical formulation is as follows:
 \begin{equation}
 	minimize\ R_0 ,
 \end{equation}
 \begin{equation}
 	s.t.\left\{ \begin{array}{c}
 		\sqrt{\left( x_i-x_j \right) ^2+\left( y_i-y_j \right) ^2}\geq 2r,\quad \forall i\ne j\\
 		\sqrt{x_{i}^{2}+y_{i}^{2}}\leq R_0-r,\quad \forall i\in \{1,...,n\}\\
 	\end{array} \right. .
 \end{equation}
Early heuristic algorithms were based on empirical or intuitive rules, aiming to find a satisfactory solution within acceptable time, often at the expense of guaranteed optimality.

In 2002, Huang et al.\cite{WU2002341} proposed a heuristic algorithm based on human experience, namely Less Flexibility First(LFF). The core idea of this algorithm is to prioritize filling the corners with the lowest flexibility, followed by the edges with moderate flexibility, and  the central area with the highest flexibility finally. By adhering to the principle of prioritizing less flexible regions, compared with the traditional random algorithms proposed earlier, this algorithm improves both packing density and computational speed. In 2005, Mladenović et al.\cite{MLADENOVIC20052419} formulated the close circle packing as a non-convex optimization problem. They proposed a heuristic algorithm based on reconstruction descent, which primarily undermines the stability of the current solution by switching coordinate systems (e.g., Cartesian coordinates and polar coordinates) to improve solution quality. This algorithm addresses the flaw of traditional methods that tend to get trapped in local optima, enhances global search capability. From 2006 to 2008, Huang et al.\cite{HUANG20062125,LU20081742} further extended equal-circle packing to unequal-circle packing. They proposed the A1.0 algorithm based on maximal hole degree and then enhanced it by incorporating a self look-ahead strategy, resulting in the A1.5 algorithm. Subsequently, they developed a hybrid heuristic algorithm that combined the maximum cavity degree principle with the PERM strategy, which further improved both packing density and computational efficiency. 

\subsection{Quasi-Physical Simulation}

The quasi-physical simulation method was first proposed by Huang et al.\cite{Huang2011} in 2011.
It comprises three key steps intially: (1)A quasi-physical descent process, where elastic forces are employed to gradually eliminate overlaps between circles, converging to a locally optimal arrangement; (2)A quasi-physical basin-hopping process, which introduces central attractive and non-contact repulsive forces to help circles escape local optima, enabling global search; (3)Radius adjustment, where a bisection method is applied to precisely adjust the circle radius $R_0$ within a fixed container, ensuring an optimal non-overlapping packing is identified. Across test instances with the number of circles $n$ ranging from 1 to 150, the algorithm improved the best-known packing solutions for 37 cases and reproduced the best-known results in 113 others. By integrating quasi-physical methods with global optimization, the approach effectively avoids local optima and enhances solution quality, highlighting the promise of quasi-physical algorithms in optimization. However, a limitation is its dependence on initial configuration, which can significantly impact convergence speed and results. Therefore, Zhu\cite{ZHU2016506}introduced a Quasi-human seniority-order(QS) algorithm to optimize initial configurations for circle packing. Its core idea is a largest-first strategy: placing larger circles first and then filling the interstitial gaps with smaller ones to maximize space utilization. The QS algorithm was combined with the quasi-physical method and benchmarked against methods employing random initial configurations. Results demonstrated that the QS algorithm significantly reduces the number of iterations and computational time, while also mitigating the risk of premature convergence or infinite loops. To solve these problems, He et al. \cite{HE201826} proposed an efficient Quasi-Physical Quasi-Human (QPQH) hybrid algorithm for equal-circle packing in 2018. Combining a modified Broyden-Fletcher-Goldfarb-Shanno(BFGS) algorithm with a quasi-human basin-hopping strategy, QPQH leverages neighborhood information of each circle for acceleration and container contraction to escape local optima. It achieved state-of-the-art results on multiple benchmarks, demonstrating high efficiency and robustness for large-scale problems.

\subsection{Metaheuristic Algorithm}

In recent years, a variety of novel metaheuristic algorithms\cite{Carrabs2014ATS,Miyazawa2016,ORICK201713,TorresEscobar2020} have been proposed for circle packing or optimal circle coverage  problems. Mainstream approaches now involve improved swarm intelligence algorithms and heuristic algorithms that incorporate classical geometric structures, such as Voronoi diagrams.

Early research tended to integrate multiple techniques to construct hybrid optimization frameworks. Ryu et al.\cite{RYU2020125076} introduced a real-time circle packing algorithm based on Voronoi diagrams in 2020. The VOROPACK-D algorithm adopts a big-circle-first strategy, modeling the gaps between placed circles via Voronoi diagrams to rapidly identify optimal positions for subsequent circles. To mitigate the impact of initial configurations on the final result, they applied a Shrink-and-Shake (S\&S) improved algorithm to  optimize the initial packing layout locally, further reducing the container radius. This approach reduced the time complexity from \(O(n^2)\) to \(O(n \log n)\). VOROPACK-D algorithm demonstrated significantly faster solution speeds than classical algorithms\cite{LOPEZ20131276,Al-Mudahka01112011}, with an average deviation from the optimum between 2\% and 5\%. However, the application of the S\&S enhancement algorithm leads to a significant increase in computational time, particularly for large-scale packing problems. 

In 2021, Chowdhury et al.\cite{CHOWDHURY2021102660} proposed a Voronoi-Glowworm Swarm Optimization-K-means (VGSOK) algorithm, which integrates K-means clustering, Voronoi diagram spatial modeling, and Glowworm Swarm Optimization. The algorithm partitions circle centers into dense regions via K-means, regarding the cluster centers as generators to construct a Voronoi diagram. It then models each circle center as a “glowworm” with an associated luciferin intensity, driving them to converge toward their corresponding Voronoi generators.  VGSOK effectively eliminates coverage holes caused by random deployment, improves coverage efficiency by 3–10\%, reduces practical energy consumption, and addresses multi‑objective optimization requirements.

As application scenarios demand greater dynamic adaptability and convergence speed, research gradually focus on improving the internal search mechanisms of metaheuristics. To address issues such as local optima entrapment (leading to "coverage holes") and slow convergence in large-scale conditions, Wang et al. \cite{biomimetics10110750} proposed a Multi-strategy Improved Flamingo Search Algorithm (MIFSA) in 2025. MIFSA incorporates several key strategies: (1) an elite opposition-based learning strategy during initialization to broaden search scope; (2) a nonlinearly decaying factor for staged step-size control, balancing global exploration and local exploitation; (3) a chaotic cosine mutation factor in later iterations to enhance population diversity and escape local optima; (4) an adaptive Lévy flight mechanism in the final stage to refine solution quality via stochastic perturbation. Experimental tests within a 50×50 square area showed that MIFSA outperformed the original flamingo search algorithm, achieving 7.48\% higher coverage at 100 iterations and 5.68\% higher at 200 iterations, with more uniform circle distribution and significantly fewer coverage holes. 

To further tackle inefficiencies and coverage holes caused by uneven circle distribution, Yang et al. \cite{Yang2025} proposed an Improved Cuckoo Search algorithm with Multiple Strategies (ICS-MS) in 2025. ICS-MS employs: (1) a phased dimension-by-dimension update strategy (full-dimensional early, dimension-wise later) to reduce computational complexity; (2) an adaptive discovery probability mechanism, dynamically adjusted based on population fitness variance; (3) bidirectional search and global-best guidance to accelerate convergence while maintaining population diversity; (4) opposition-based learning on elite individuals for a subsequent deep search in their vicinity. Compared against metaheuristic algorithms under various circle counts (20/30) and iterations (200/1000), ICS-MS improved coverage rates by 2.32\% to 22.21\% and yielded more uniform circles distributions.

Meanwhile, Shaikh et al.\cite{Shaikh2025} introduced an Improved Chaotic Grey Wolf Optimizer (ICGWO) in 2025. To mitigate the traditional Grey Wolf Optimizer's tendencies toward local optima and slow convergence, ICGWO integrates Gaussian chaotic mapping into the position-updating process, leveraging chaotic sequences to diversify search paths. Unlike comparative algorithms (e.g., Particle Swarm Optimization or Ant Colony Optimization) that require intricate parameter tuning, ICGWO maintains a balance between global exploration and local exploitation with minimal parametric overhead. When compared on identical circle counts, ICGWO surpasses existing counterparts by 2.18–16.41\% in coverage rate, with pronounced advantages in large-scale deployments. The algorithm demonstrates high scalability and robustness, which is suitable for complex real-world scenarios.

 \section{Preliminaries}
 \subsection{Convex Set and Convex Hull}
  This paper focuses on the covering of convex polygons, a concept originating from convex sets and convex hulls.
 
 \begin{definition}[Convex Set]
  Given a set \( S \), if for any \( \forall x, y \in S \), \( (1-\lambda)x + \lambda y \in S \) holds for all \( \forall \lambda \in [0,1] \), then \( S \) is a convex set. Geometrically, this intuitively means that a convex set contains the line segment connecting any two points within it.
 \end{definition}

\begin{definition}[Convex Hull]
 The convex hull \(\text{Conv}(S)\) of a set S is the smallest convex set containing S, that is, the intersection of all convex sets that contain S.
\end{definition}

A convex polygon in the plane can be defined as the boundary of the convex hull of a finite set of points~$S$. Such a polygon is typically represented by starting at an arbitrary vertex of the convex hull and connecting all vertices sequentially in clockwise (or counterclockwise) order to form a closed region. Consequently, the problem of constructing a convex polygon reduces to computing the convex hull of its given point set~$S$. Well-known algorithms for computing the convex hull include Graham Scan~\cite{GRAHAM1972132}, Jarvis March~\cite{JARVIS197318}, and Monotone Chain~\cite{ANDREW1979216}.

\begin{lemma}
\cite{berg2008}. Given any planar point set containing n points, its convex hull can be constructed in \(O(n\log n)\) time.
\end{lemma}

\subsection{Voronoi Diagram}

A Voronoi diagram is a partition of a planar region that divides the plane into \( n \) cells. Given a set of points \( S = \{s_1, s_2, \cdots, s_n\} \), these \( n \) distinct points are referred to as generators, with each generator corresponding to a partitioned cell. For example, the cell corresponding to generator \( s_i \) is denoted $\mathcal{V}(s_i)$, and the Voronoi diagram associated with point set \( S \) is denoted \( Vor(S) \).

\begin{lemma}
	\cite{berg2008}. A point $p \in \mathcal{V}(s_i)$ if and only if \( \delta(p, s_i) < \delta(p, s_j) \) for all \( s_j \in S (j \neq i) \), where \( \delta(p, s_i) \) represents the Euclidean distance between \( p \) and \( s_i \). 
\end{lemma}

\begin{figure}[htbp]
	\centering
	\begin{subfigure}[b]{0.35\textwidth}
		\centering
		\includegraphics[width=\textwidth]{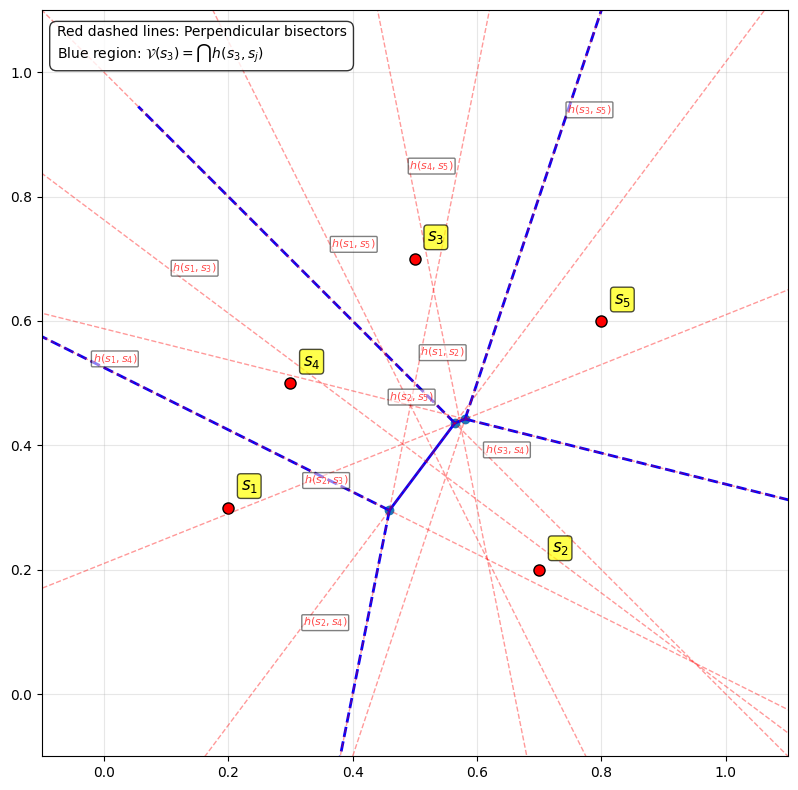}
		\caption{Intersection of all half-planes}
		\label{fig:half_planes}
	\end{subfigure}
	\begin{subfigure}[b]{0.35\textwidth}
		\centering
		\includegraphics[width=\textwidth]{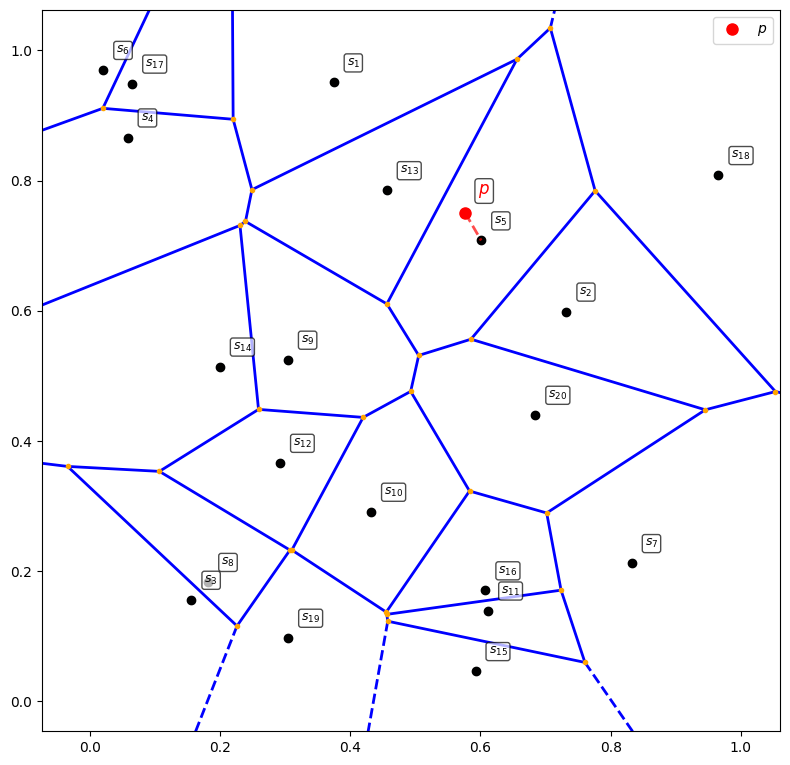}
		\caption{Voronoi diagram in $\mathbb{R}^2$}
		\label{fig:voronoi}
	\end{subfigure}
	\caption{Construction of Voronoi diagrams through half-plane intersections}
	\label{fig:1}
\end{figure}

For any two distinct points \( s_i \) and \( s_j \) in the plane, the perpendicular bisector of the line segment \( \overline{s_i s_j} \) partitions the plane into two half-planes. The open half-plane containing \( s_i \) is denoted by \( h(s_i, s_j) \), while the open half-plane containing \( s_j \) is denoted by \( h(s_j, s_i) \).It follows that the Voronoi cell of \( s_i \) is given by the intersection of all such half-planes induced by other generators:
\begin{equation}
\mathcal{V}(s_i) = \bigcap_{\substack{1 \le j \le n \\ j \ne i}} h(s_i, s_j).
\end{equation}
The Voronoi diagram of the set \( S \) is then the union of all Voronoi cells:
\begin{equation}
Vor(S) = \bigcup_{i=1}^n \mathcal{V}(s_i).
\end{equation}
This construction is illustrated in Fig.~\ref{fig:1}.

If all points of $S$ are collinear, then $Vor(S)$ consists of $n-1$ parallel lines. Otherwise, $\operatorname{Vor}(S)$ is connected, and each of its edges is either a line segment or a ray.

\begin{lemma}
	\cite{berg2008}. If $n \geq 3$, the Voronoi diagram of $n$ generators has at most $3n - 6$ edges and at most $2n - 5$ vertices. 
\end{lemma}

\begin{definition}[Largest Empty Circle]
 For any point $q$ in $\operatorname{Vor}(S)$, the largest circle centered at $q$ that contains no generators of $S$ in its interior is called the largest empty circle of $q$ with respect to $S$, denoted by $C_S(q)$.
\end{definition}

\begin{theorem}\label{thm:voronoi_vertex_edge}
 	\cite{berg2008}. For the Voronoi diagram $Vor(S)$ of an arbitrary point set $S$, the following properties hold:
	\begin{enumerate}
		\item[(1)] A point $q$ is a vertex of $\operatorname{Vor}(S)$ if and only if there exist at least three generators on the boundary of its largest empty circle $C_S(q)$.
		\item[(2)] The bisector between $s_i$ and $s_j$ defines an edge in $\operatorname{Vor}(S)$ if and only if there exists a point $q$ on this bisector such that the boundary of $C_S(q)$ passes through $s_i$ and $s_j$, but contains no other generators.
	\end{enumerate}
\end{theorem}

\begin{figure}[H]
 	\centering
 	\begin{minipage}[b]{0.4\textwidth}
 		\centering
 		\includegraphics[width=\textwidth]{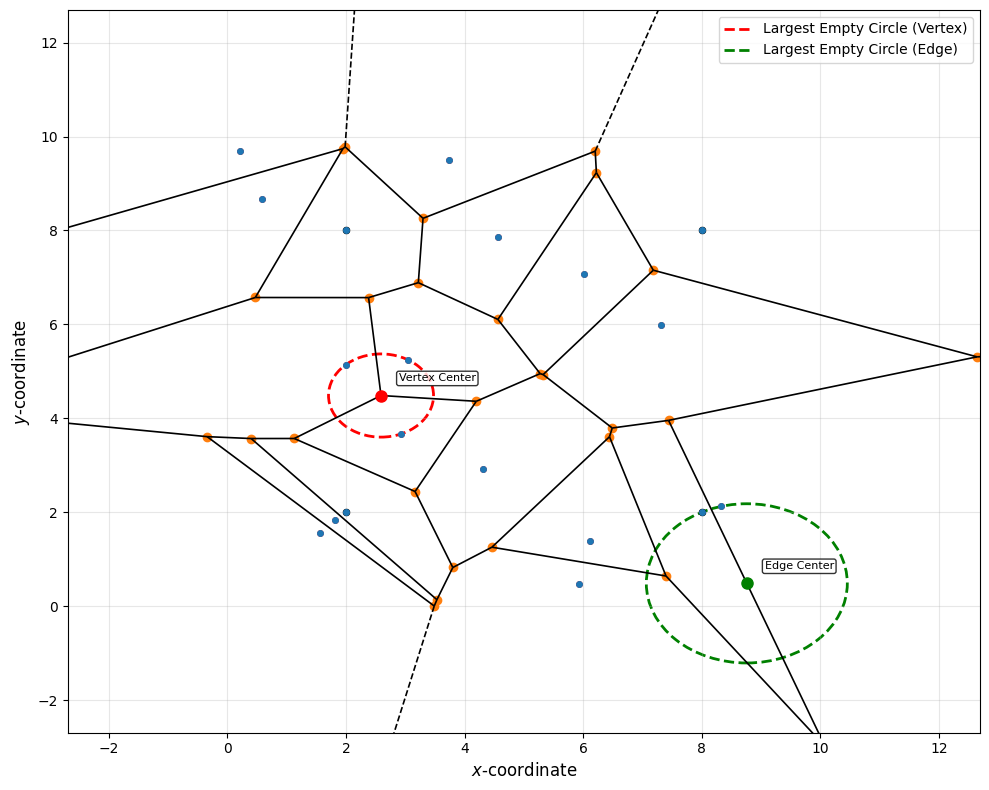}
 		\caption{Two Types of Largest Empty Circle}
 		\label{fig:lec_types}
 	\end{minipage}
 	\begin{minipage}[b]{0.42\textwidth}
 		\centering
 		\includegraphics[width=\textwidth]{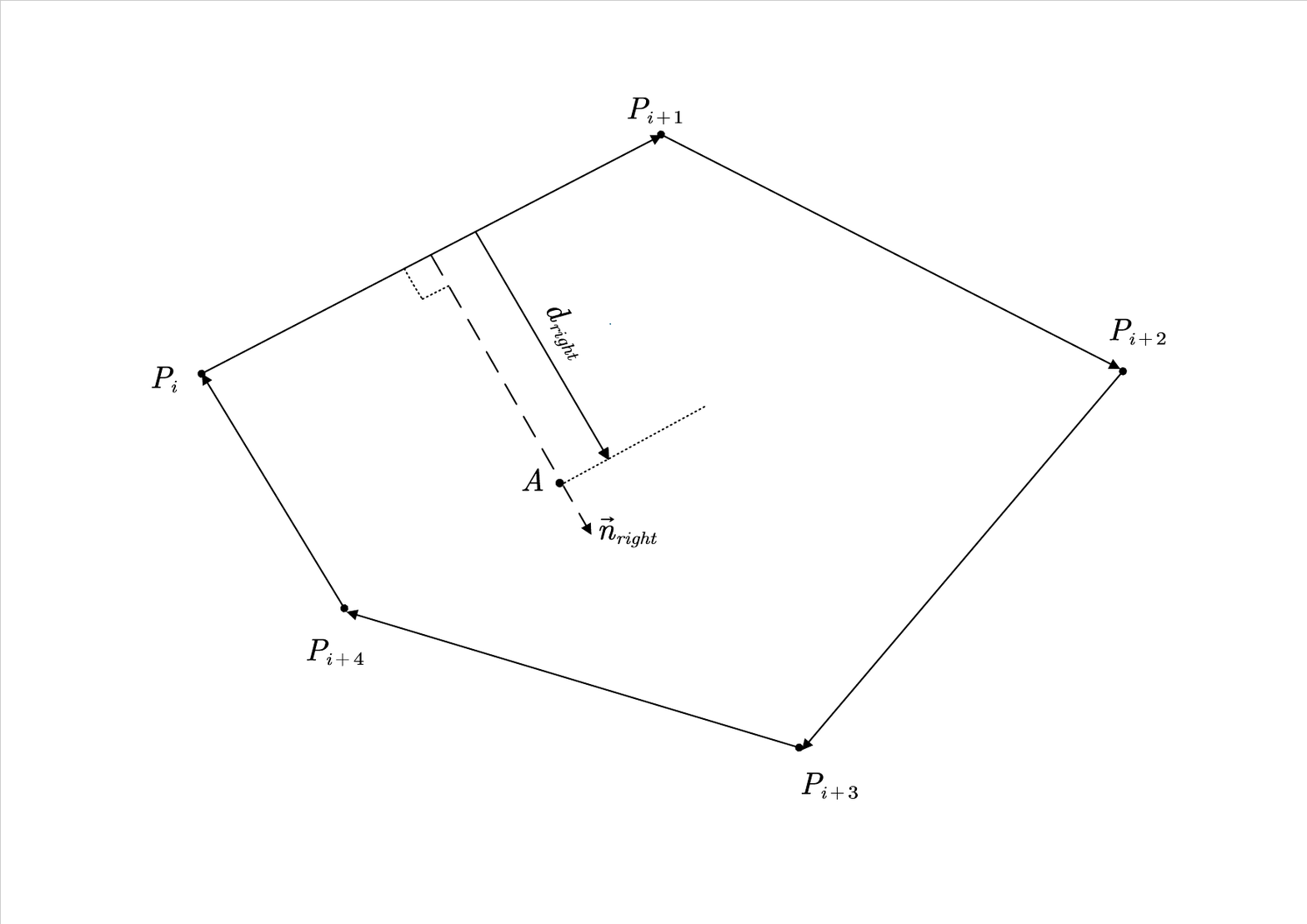}
 		\caption{A convex polygon composed of clockwise-oriented directed line segments}
 		\label{fig:polygon_diagram}
 	\end{minipage}
 	\vspace{5pt} 
 \end{figure}

As shown in Theorem~\ref{thm:voronoi_vertex_edge} and  Fig.~\ref{fig:lec_types}, the vertices and edges of Voronoi diagrams have clear geometric interpretations.
 
 Voronoi diagrams are typically constructed by using  a divide-and-conquer approach \cite{book2000} or indirectly via Delaunay triangulation, leveraging their geometric duality
 \cite{preparata2012computational}. We employ this framework to assess distribution uniformity by constructing the Voronoi diagram of circle centers, as detailed in Section 4.2.

 \subsection{Geometric Computation}
 This paper focuses on convex polygons, thus it is necessary to verify the convexity of arbitrary polygons. The primary method is the "rotation method" \cite{SCHORN19947}, which treats the polygon boundary as a closed sequence of consistently oriented directed line segments, as illustrated in Fig.~\ref{fig:polygon_diagram}.
 
 \begin{lemma}
 \cite{SCHORN19947}. For each vertex $P_i = (x_i, y_i)$ of the polygon, compute the cross product of vectors formed by three consecutive vertices $P_i \to P_{i+1} \to P_{i+2}$ sequentially. If all cross products have the same sign (either all positive or all negative), then the polygon is convex. Specifically, let $\overrightarrow{P_iP_{i+1}} = (x_{i+1} - x_i, y_{i+1} - y_i)$
 and $\overrightarrow{P_{i+1}P_{i+2}} = (x_{i+2} - x_{i+1}, y_{i+2} - y_{i+1}),$
 then the cross product $\overrightarrow{P_iP_{i+1}} \times \overrightarrow{P_{i+1}P_{i+2}} \quad (i = 1, \dots, n-2)$
 must be consistently signed.
 \end{lemma}

To determine whether the center of a circle lies inside a convex polygon, it is necessary to introduce the signed distance function. 
 
 \begin{definition}[Signed Distance Function] 
If $P$ is a subset of a metric space $X$ with metric $d$, the signed distance function $f$ is defined as follows:
$$
f_P\left( x \right) =\left\{ \begin{array}{l}
	d\left( x,\partial P \right) ,\ if\ x\in P\\
	-d\left( x,\partial P \right) ,\ if\ x\notin P\\
\end{array} \right. .
 $$
 Here, $\partial P$ denotes the boundary of $P$. For any $x \in X$, $d(x, \partial P) := \inf_{y \in \partial P} d(x, y).$
 \end{definition}
 
 For simplicity of notation and computation, let $d(x, y)$ denote the Euclidean distance between points $x$ and $y$. The concepts of the right normal vector and the right distance for a directed segment are defined as follows.
 
 \begin{definition}[Right Normal Vector] 
 Given a directed segment $\overrightarrow{P_1P_2}$, its right normal vector is the unit normal vector that points to the right-hand side when traversing from $P_1$ to $P_2$ \cite{hubbard2015vector}. Formally, for the direction vector $\vec{v} = (v_x, v_y) = \overrightarrow{P_1P_2}$, the right normal vector is given by
 $\vec{n}_{\mathrm{right}} = (v_y, -v_x).$
 \end{definition}
 
 \begin{definition}[Right Distance] 
 The right distance from a point $A$ to the directed segment $\overrightarrow{P_1P_2}$ is defined to be positive if $A$ lies on the right-hand side of the segment, negative if on the left-hand side, and zero if on the line containing the segment. It is computed by the formula:
 $d_{\mathrm{P_1P_2}}(A) =  \frac{\overrightarrow{P_1A} \cdot \vec{n}_{\mathrm{right}}}{\|\vec{n}_{right}\|}$
 	, where $\vec{n}_{\mathrm{right}}$ is the right normal vector as defined above.
 \end{definition}
 
Then we obtain Lemma~\ref{lemma:point_containment_convex_polygon}  for point containment in convex polygons.

\begin{lemma}\label{lemma:point_containment_convex_polygon}
	A point $A$ is contained within a convex polygon $P$ if and only if
$f_P(A) = \min_{j} d_{j}(A) > 0$,
where $j$ denotes the edge index of the polygon.
\end{lemma}
 
 \section{Improved Quasi-Physical Dynamic Algorithm }
 
The quasi-physical method was initially proposed by Huang et al. \cite{WENQI2002195,Huang2011} for solving the equal-circle packing problem. The core idea is to model the circles as elastic balls confined within the convex polygon boundary, which acts as a container wall. By simulating the collision and squeezing motions among these elastic balls, they gradually spread out within the polygon, leading to an efficient packing configuration. This thought can be adapted to the circle covering optimization problem in convex polygons. We further improve the method by refining the initial distribution, iterative process, and boundary covering strategy, proposing an improved quasi-physical dynamic algorithm.
 
 \subsection{Quasi-physical Elastic Circle Model}
 
 \begin{definition}[Configuration] 
The vector of circle centers at any time, denoted as $\boldsymbol{C}=\left[x_1,y_1,x_2,y_2,\cdots ,x_n,y_n\right]$, defines the configuration comprising a convex polygon and circles.
 \end{definition}
 
\begin{figure}[htbp]
 	\centering
 	\begin{subfigure}[b]{0.48\textwidth}
 		\includegraphics[width=\textwidth]{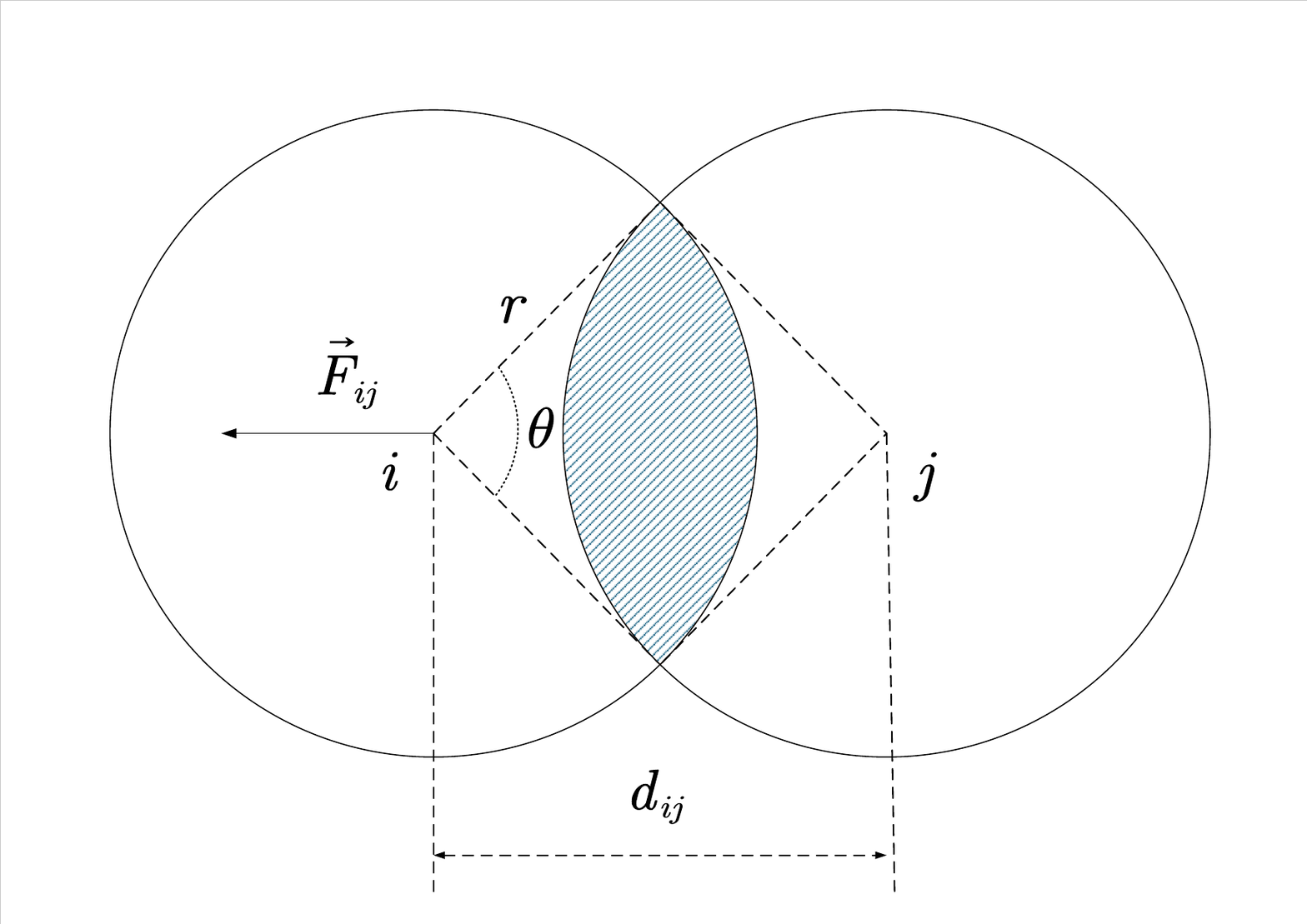}
 		\caption{ Force between circles}
 		\label{fig:left}
 	\end{subfigure}
 	\begin{subfigure}[b]{0.49\textwidth}
 		\includegraphics[width=\textwidth]{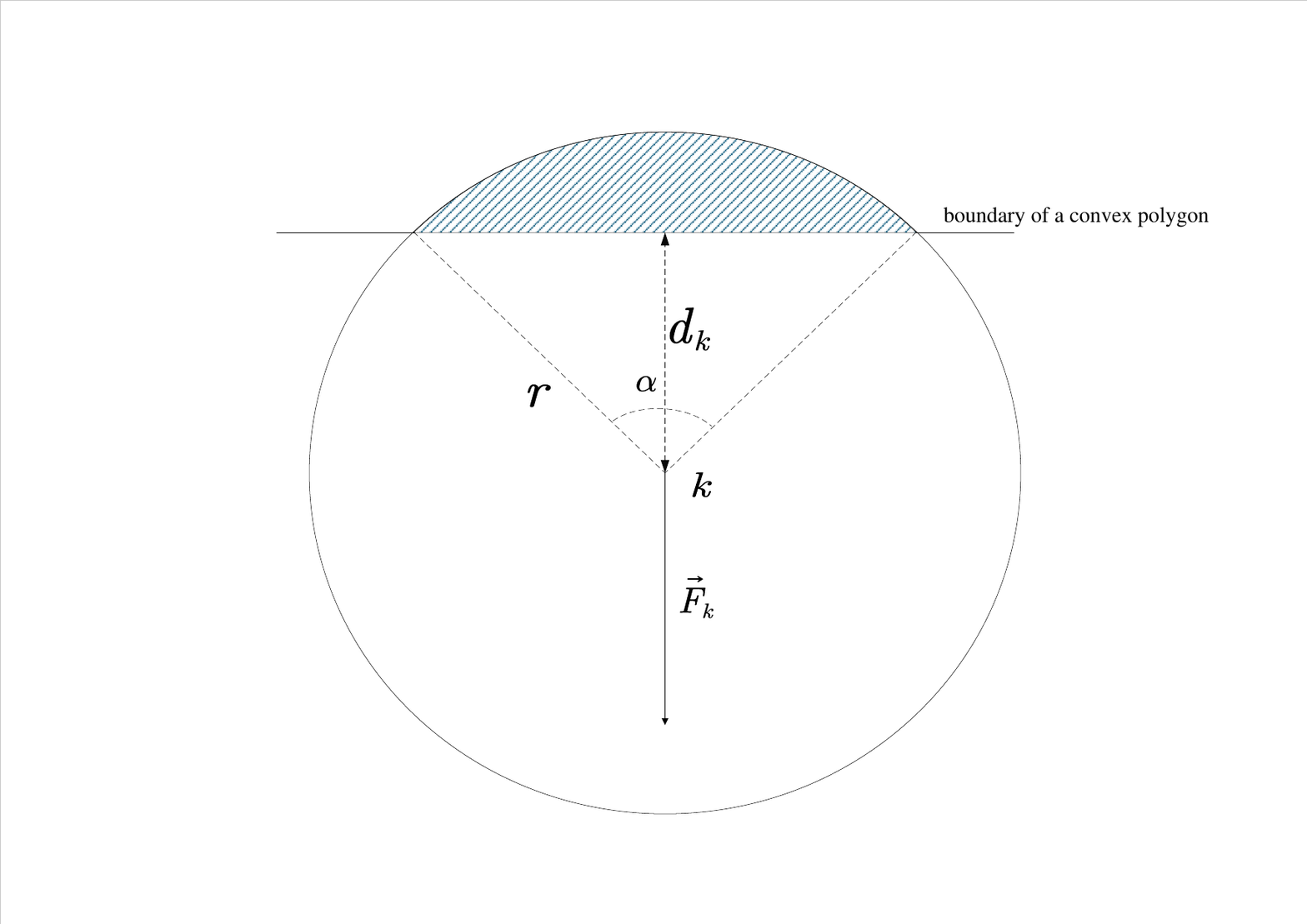}
 		\caption{Force between a circle and the  boundary}
 		\label{fig:right}
 	\end{subfigure}
 	\caption{The two types of elastic forces}
 	\label{fig:total}
 \end{figure}

 The system composed of a convex polygon and circles involves two types of elastic forces: circle–circle and circle–polygon boundary. Let \(\vec{F}_{ij}\) denote the elastic force exerted by circle \(j\) on circle \(i\), directed from \(j\) toward \(i\), and let \(\vec{d}_{ij}\) represent the displacement vector from the center \(j\) to the center \(i\). Similarly, let \(\vec{F}_k\) denote the elastic force exerted by the polygon boundary on circle \(k\), and \(\vec{d}_k\) the displacement vector from the polygon boundary to the center \(k\), as illustrated in Fig.\ref{fig:total}.
 
We model the overlapping area between a circle and a convex polygon boundary, as well as the overlapping area between two circles, as the magnitudes of two types of elastic forces (represented by the shaded areas). A larger overlapping area corresponds to a stronger force. Therefore, the shaded areas in Fig. (a) and (b), denoted as $S_a$ and $S_b$, are given by:
 \begin{equation}
 	\left\{ \begin{array}{l}
 		\theta =2\arccos \left( \frac{d_{ij}/2}{r} \right)\\
 		S_a=2\left( \frac{1}{2}r^2\theta -2r\sin \frac{\theta}{2}r\cos \frac{\theta}{2} \right) =r^2\left( \theta -\sin \theta \right)\\
 	\end{array} \right. ,
 \end{equation}
 
 \begin{equation}
 	\left\{ \begin{array}{l}
 		\alpha =2\arccos \left( \frac{d_k}{r} \right)\\
 		S_b=\frac{1}{2}r^2\alpha -2\sqrt{r^2-d_{k}^{2}}d_k\cdot \frac{1}{2}=\frac{1}{2}r^2\alpha -d_k\sqrt{r^2-d_{k}^{2}}\\
 	\end{array}. \right. 
 \end{equation}
 Therefore, the two types of elastic forces can be expressed as:
 \begin{equation}
	\left\{ \begin{array}{l}
		\vec{F}_{ij}=S_a\cdot \frac{\vec{d}_{ij}}{\left| \vec{d}_{ij} \right|}\\
		\vec{F}_k=S_b\cdot \frac{\vec{d}_k}{\left| \vec{d}_k \right|}\\
	\end{array} \right. ,
\end{equation}
 where \( S_a \) and \( S_b \) denote the overlapping areas defined previously. The direction of the elastic force between two circles always points from the center of the exerting circle to the center of the receiving circle, while the direction of the elastic force between a circle and the polygon boundary is given by the rightward normal vector of the directed edge. 
 
 Since multiple circles may overlap or a circle may intersect with multiple edges of the polygon, the total elastic force acting on circle \( i \) is given by:
 
\begin{equation}
  	\vec{F}_{\text{elastic,i}}=\sum_{j=1,j\ne i}^n{\vec{F}_{ij}}+\sum_{k=1}^N{\vec{F}_k}.
  	\label{10}
  \end{equation}

 \subsection{Structure-Preserving Initialization Strategy}
 
 The initial configuration of circles significantly influences the quality of the final near-optimal solution and computation time. Therefore, a well-chosen initial configuration is essential for subsequent high-quality optimization iterations. Inspired by the close packing principle of regular hexagons\cite{toth1949dichteste}, we propose a structure-preserving initialization strategy based on hexagonal close packing.
 
\begin{figure}[htbp]
	\centering
	\includegraphics[width=0.55\textwidth]{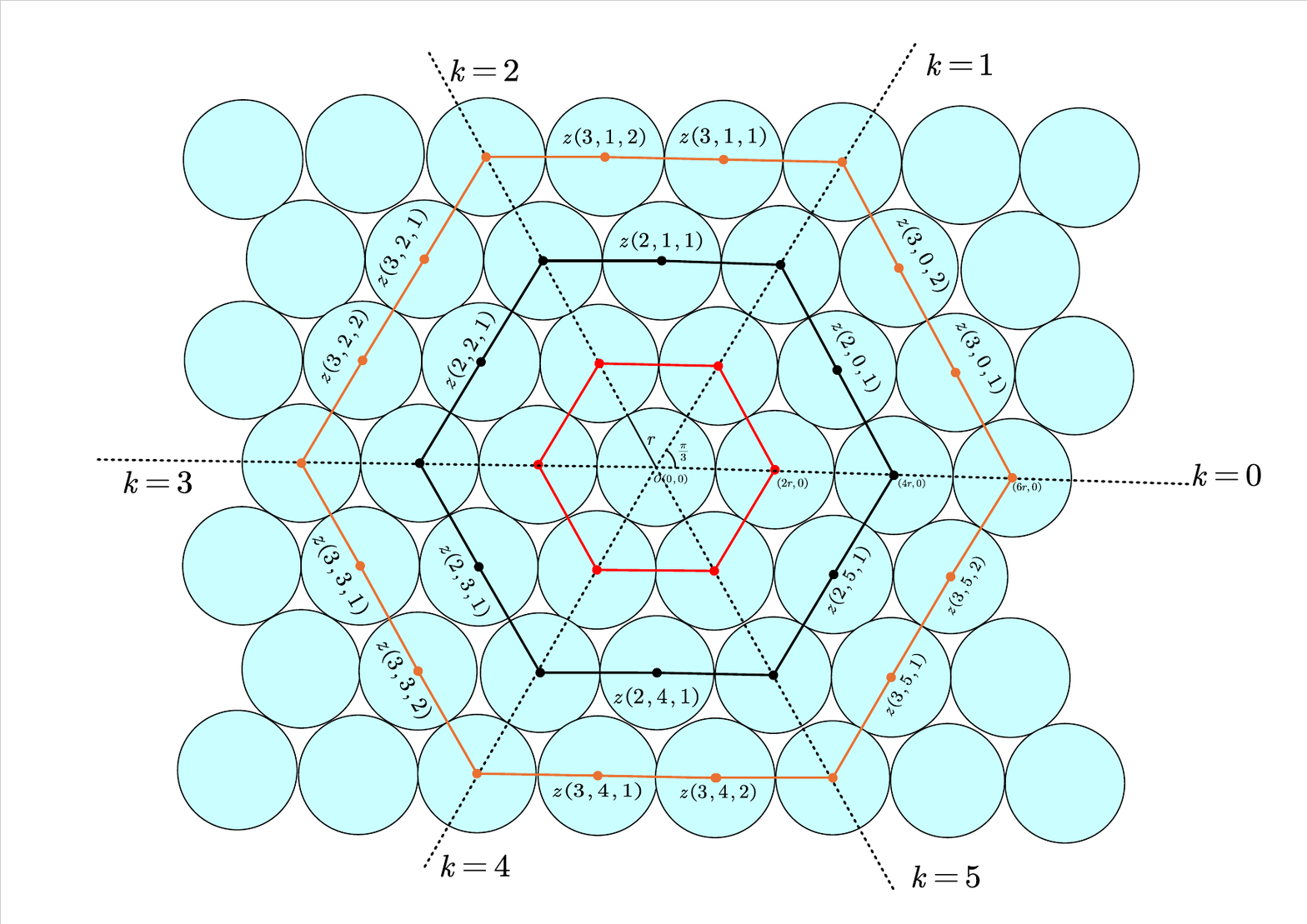}
	\caption{Regular hexagonal close packing of circles generated from the center outward.}
	\label{Figure:5}
\end{figure}

\begin{figure}[htbp]
 	\centering
 	\includegraphics[width=0.8\textwidth]{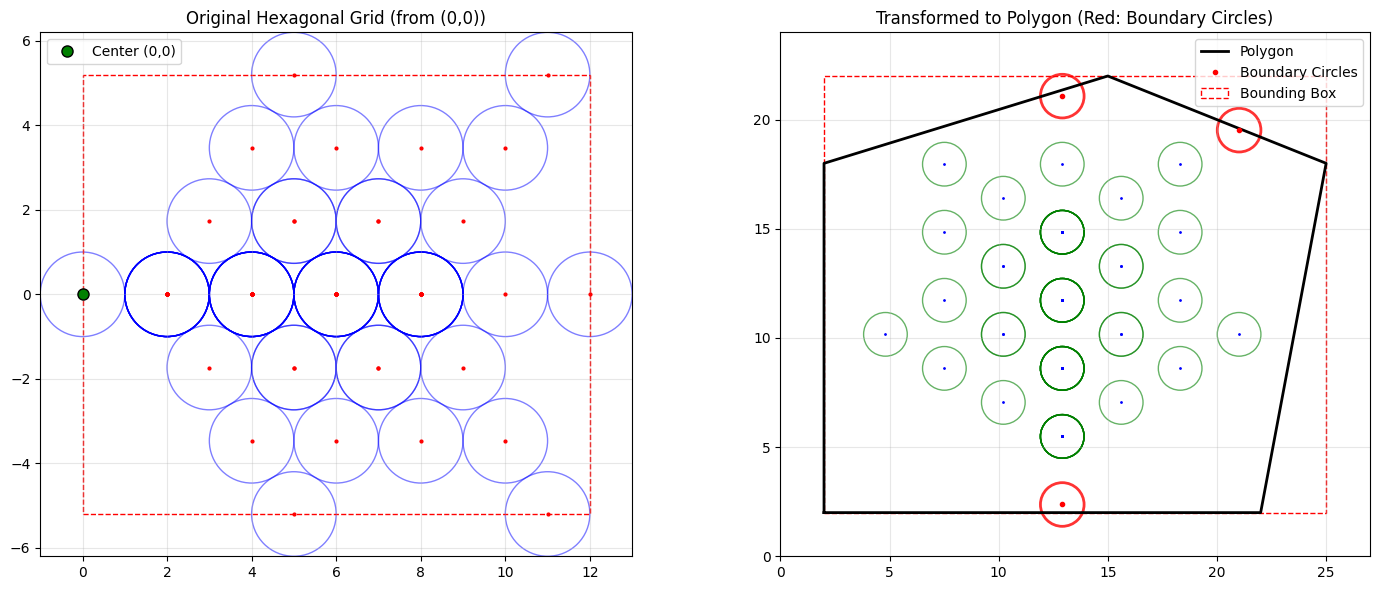} 
 	\caption{The projection of densely packed circles into a convex polygonal domain through scaling and affine transformations}
 	\label{figure:6}
 \end{figure}

 First, a series of concentric regular hexagons is generated outward from the center on a base plane. This configuration is then projected into the target convex polygon through scaling and affine transformation, directly transferring the close-packed circle pattern into the polygonal domain, as shown in Fig.~\ref{Figure:5} and~\ref{figure:6}.
 
 This approach ensures that the circles in the interior region of the polygon are already in a near-optimal close-packed arrangement, requiring no further iterative optimization. Consequently, the subsequent optimization process can focus exclusively on adjusting the circles near the boundary. This strategy not only preserves the optimal packing structure across most of the area but also significantly reduces the computational cost of iterative optimization, leading to a substantial improvement in overall algorithmic efficiency.
 
 \begin{definition} [Circular Sector]
 A circular sector is defined as a planar region bounded by two radii emanating from the circle's center and the arc between them. According to Fig.~\ref{Figure:5},We denote the region between $k=i$ and $k=i+1$ as the i-th sector.
 \end{definition}
 
Assume that the 0-th layer consists of a single circle with its center at \((0, 0)\). The first layer is composed of six circles generated around the initial circle along six respective directions, as represented in Fig. ~\ref{Figure:5}. This pattern continues recursively: for the \(l\)-th layer (\(l \geq 1\)), \(l\) circles are generated along each of the six directions (with an angular interval of \(\frac{\pi}{3}\) radians between adjacent directions). The distance from the center of each circle in this layer to the origin is \(2lr\), and the distance between the centers of any two adjacent circles is \(2r\). To facilitate a concise formulation, complex numbers are employed for the circle center coordinates. Given a rotation angle of \(\theta = k \cdot \frac{\pi}{3}\) (where \(k = 0, 1, \dots, 5\)), the complex coordinate of the \(m\)-th circle in the \(k\)-th sector of the \(l\)-th layer is
 \begin{equation}
	z\left( l,k,m \right) =2r\cdot l\cdot e^{i\cdot k\cdot \frac{\pi}{3}}+2r\cdot m\cdot e^{i\cdot \left( k+2 \right) \cdot \frac{\pi}{3}},
	\label{eq:11}
\end{equation}
where $ k=0,\cdots ,5,m=0,\cdots ,l-1$.
 
\begin{definition}[Minimum Bounding Rectangle, MBR]
 The Minimum Bounding Rectangle of a polygon is defined as the rectangle of the smallest area that can fully contain the polygon. For any convex polygon, the orientation of its MBR is commonly used to determine the polygon's principal direction.
\end{definition}
 
\begin{figure}[H]
 	\centering
 	\includegraphics[width=0.85\textwidth]{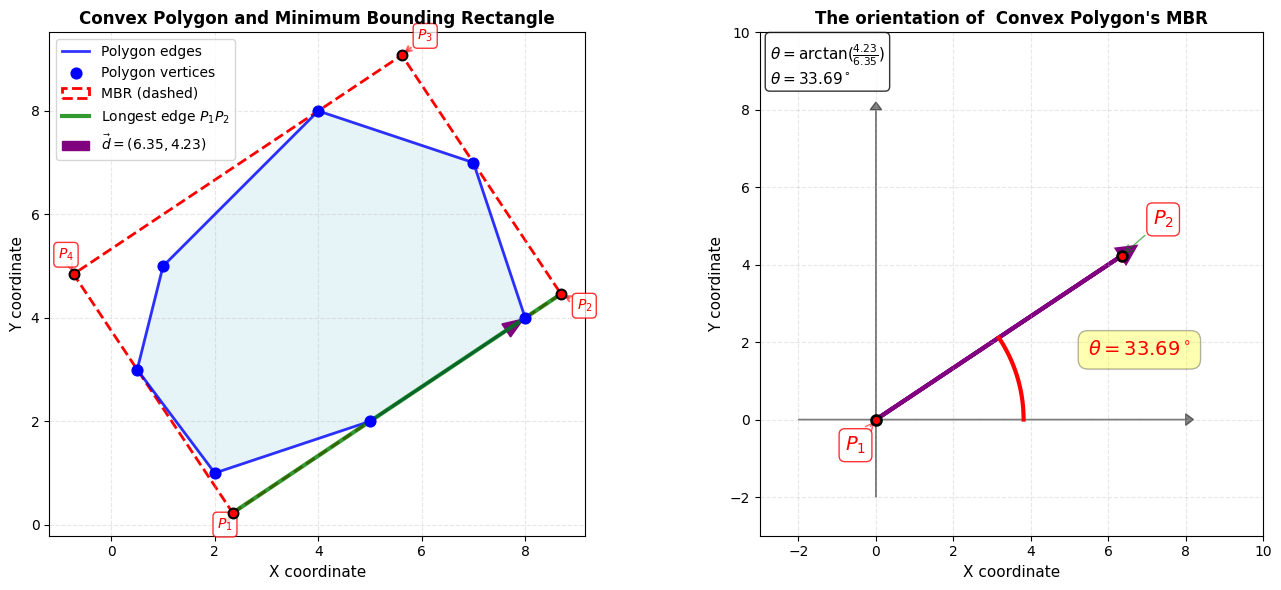} 
 	\caption{Convex Polygon $P$ and its MBR}
 	\label{figure:7}
 \end{figure}

 Suppose the vertex coordinates of minimum bounding rectangle are denoted as \(P_1=(x_1,y_1), P_2=(x_2,y_2), P_3=(x_3,y_3), P_4=(x_4,y_4)\) (arranged in a clockwise order). Without loss of generality, let \(P_1P_2\) be the longest side of the rectangle. Then, the direction vector of the rectangle is given by \(\vec{d} = (x_2 - x_1, y_2 - y_1)\), and the corresponding rotation angle is calculated as \(\theta = \arctan\left(\frac{y_2 - y_1}{x_2 - x_1}\right)\). For instance, consider an arbitrary random convex polygon \(P = \{(0.5, 3), (2, 1), (5, 2), (8, 4), (7, 7), (4, 8), (1, 5)\}\). Its principal direction is illustrated in Fig.\ref{figure:7}.

Based on the orientation analysis of the minimum bounding rectangle, we propose an initialization algorithm with structure-preserving strategy for close-packed hexagonal circles. The main steps are as follows:

 \textbf{Step 1:} Compute the minimum bounding rectangle of an arbitrary convex polygon using the rotating calipers algorithm \cite{carlos_2020,h._2008};

 \textbf{Step 2:} Calculate the direction vector $\vec{d}$ and the rotation angle $\theta$ of the rectangle;
 
 \textbf{Step 3:} Align the orientation of the hexagonal lattice with the principal direction of the polygon, as follow:
 \begin{equation}
 	z_{rolated}\left( l,k,m \right) =z\left( l,k,m \right) \cdot e^{i\theta};
 \end{equation}
 
 \textbf{Step 4:} Calculate the scaling factor $\beta$ to scale the hexagonal lattice to match the polygon's size:
 \begin{equation}
 	\left\{ \begin{array}{l}
 		\beta =\min \left( \frac{w_p-2r}{w_h},\frac{h_p-2r}{h_h} \right) \cdot \alpha\\
 		z_{scaled}\left( l,k,m \right) =\beta \cdot z_{rotated}\left( l,k,m \right)\\
 	\end{array} \right. ,
 \end{equation}
 Here, $w_p$ and $h_p$ denote the length and width of the polygon's minimum bounding rectangle, respectively, while $w_h$ and $h_h$ represent the length and width of the bounding box of the hexagonal lattice. $r$ is the radius of the circles. To prevent the transformed circles from overflowing the polygon boundary, we define $\alpha \in (0.9, 0.95)$ as a safety coefficient. The terms $\frac{w_p-2r}{w_h}$ and $\frac{h_p-2r}{h_h}$ represent the ratios of the available space within the polygon to the original sizes of the hexagonal lattice along the length and width directions, respectively. The safety coefficient $\alpha$ is introduced to account for numerical computation errors and the irregular shape of the polygon, thereby reserving adjustment space for subsequent optimization iterations.
 
 \textbf{Step 5:} Place the transformed hexagonal lattice within the polygon by aligning the lattice center with the polygon centroid $\boldsymbol{c}_p$, resulting in a roughly centered initialization as illustrated in Fig.~\ref{figure:6}:
 
 \begin{equation}
 	\left\{ \begin{array}{l}
 		\boldsymbol{c}_p=\frac{1}{N}\sum_{i=1}^N{\boldsymbol{v}_i}=\left( \frac{1}{N}\sum{x_i},\frac{1}{N}\sum{y_i} \right)\\
 		z_{final}\left( l,k,m \right) =z_{scaled}\left( l,k,m \right) +\boldsymbol{c}_p\\
 	\end{array} \right. ,
 \end{equation}
where $\boldsymbol{v}_i = (x_i, y_i)$ denotes the polygon vertices and $N$ is the number of vertices.
 
 \textbf{Step 6:} Perform boundary detection on the affine-transformed circles and filter out those exceeding the polygon boundary using the boundary-aware filtering algorithm \cite{xianguang_2024}. Subsequently, apply the corner-occupancy principle \cite{WU2002341} to insert circles near the convex polygon interior.

Initialization algorithm enables the circles to maintain hexagonal close-packing characteristics within the polygon. The distance between adjacent circle centers becomes \(2\beta r\), while hexagonal symmetry is preserved. When the number of circles is small, the initial distribution corresponds to the optimal configuration, thus transforming the maximum area coverage problem into a traditional packing problem. Even with a larger number of circles, the algorithm establishes a high-quality starting point for subsequent iterative optimization, significantly improving computational efficiency.
 
\begin{algorithm}[H]
	\caption{Structure-Preserving Initialization Strategy}
	\label{alg:lattice_initialization}
	\begin{algorithmic}[1]
		\Require Convex polygon $P$, circles'number $N$, radius $r$, safety coefficient $\alpha \in (0.9, 0.95)$
		\Ensure Initial circle positions $\boldsymbol{C}^{(0)} = \{\boldsymbol{c}_1^{(0)}, \dots, \boldsymbol{c}_N^{(0)}\}$
		
		\State \textbf{Step 1: Compute minimum bounding rectangle}
		\State $\text{Rect}, \vec{d}, \theta \gets \text{RotatingCalipers}(P)$ 
		\State $w_p, h_p \gets \text{width}(\text{Rect}), \text{height}(\text{Rect})$
		
		\State \textbf{Step 2: Generate base hexagonal lattice}
		\State $\mathcal{L}_{\text{base}} \gets \text{GenerateHexagonalLattice}(N)$ according to equation \eqref{eq:11}
		\State $w_h, h_h \gets \text{BoundingBoxDimensions}(\mathcal{L}_{\text{base}})$
		
		\State \textbf{Step 3: Rotate lattice}
		\State $\mathcal{L}_{\text{rot}} \gets \{ z \cdot e^{i\theta} \mid z \in \mathcal{L}_{\text{base}} \}$
		
		\State \textbf{Step 4: Scale lattice}
		\State $\beta \gets \min\left( \frac{w_p - 2r}{w_h}, \frac{h_p - 2r}{h_h} \right) \cdot \alpha$
		\State $\mathcal{L}_{\text{scaled}} \gets \{ \beta \cdot z \mid z \in \mathcal{L}_{\text{rot}} \}$
		\State \textbf{Step 5: Align lattice center with polygon centroid}
		\State $\boldsymbol{c}_p \gets \frac{1}{|V(P)|} \sum_{\boldsymbol{v} \in V(P)} \boldsymbol{v}$ 
		\State $\mathcal{L}_{\text{trans}} \gets \{ z + \boldsymbol{c}_p \mid z \in \mathcal{L}_{\text{scaled}} \}$
		\State \textbf{Step 6: Filter and optimize}
		\State $\mathcal{L}_{\text{filtered}} \gets \text{BoundaryAwareFilter}(\mathcal{L}_{\text{trans}}, P)$ 
		\State $\mathcal{L}_{\text{final}} \gets \text{CornerOccupancyInsertion}(\mathcal{L}_{\text{filtered}}, P)$ 
		\State \Return $\mathcal{L}_{\text{final}}$
	\end{algorithmic}
\end{algorithm}

\subsection{Virtual-Force and Radius Iteration Process}
After initializing the coordinates of all circle centers, we perform dynamic modeling of the system consisting of all circles and the convex polygon.  Assume each circle \( C_i \) is a dynamic object with three attributes: position \( \vec{x}_i = (x_i, y_i) \), velocity \( \vec{v}_i = (v_{ix}, v_{iy}) \), and the resultant external force $\vec{F}_{\text{total},i}=(\vec{F}_{\text{total},ix},\vec{F}_{\text{total},iy})$. 

To prevent the circles from moving indefinitely due to inertia, we introduce a viscous damping term \( \vec{F}_{\text{friction}} \) for each circle.  
Let the viscous coefficient be \( \mu \); the damping force acts opposite to the velocity direction and is proportional to the magnitude of the velocity.  
Thus, the resultant external force on each circle is
 \begin{equation}
 	\left\{ \begin{array}{l}
 		\vec{F}_{\text{friction}}=-\mu\cdot \vec{v}\\
 		\vec{F}_{\text{total,i}}=\sum_{j=1,j\ne i}^n{\vec{F}_{ij}}+\sum_{k=1}^N{\vec{F}_k}+\vec{F}_{\text{friction}}\\
 	\end{array} \right. .
 \end{equation}
 
To simplify the computation, the mass of each circle can be set to 1 and the time step to $\Delta t$. According to Newton's second law, the update equations for the acceleration, velocity, and position of each circle are then given by:
 \begin{equation}
 	\left\{ \begin{array}{l}
 		\vec{a}_i=\vec{F}_{total,i}\\
 		\vec{v}_{i}^{\left( t+\varDelta t \right)}=\vec{v}_{i}^{\left( t \right)}+\vec{a}_i\varDelta t\\
 		\vec{x}_{i}^{\left( t+\varDelta t \right)}=\vec{x}_{i}^{\left( t \right)}+\vec{v}_{i}^{\left( t \right)}\varDelta t\\
 	\end{array} \right. .
 \end{equation}
In the Python program, updates for velocity and position are performed separately along the $x-$ and $y-$ axes, i.e,
 \begin{equation}
 	\left\{ \begin{array}{l}
 		v_{x}^{\left( t+\varDelta t \right)}=v_{x}^{\left( t \right)}+a_x\varDelta t\\
 		v_{y}^{\left( t+\varDelta t \right)}=v_{y}^{\left( t \right)}+a_y\varDelta t\\
 	\end{array} \right. 
 \end{equation}
 
 \begin{equation}
 	\left\{ \begin{array}{l}
 		x^{\left( t+\varDelta t \right)}=x^{\left( t \right)}+v_x\varDelta t\\
 		y^{\left( t+\varDelta t \right)}=y^{\left( t \right)}+v_y\varDelta t\\
 	\end{array} \right. 
 \end{equation}
 
 Upon completion of the dynamical modeling, we employ a progressive radius expansion strategy to optimize the distribution of circles. This strategy alternates between radius expansion and virtual-force balancing to drive the system from an initial relaxed state toward the target configuration. Specifically, the algorithm starts with a set of radii smaller than the target values. In each iteration, it first increases all radii by a predefined step size $\Delta r_k$, followed by a virtual-force simulation that brings the circles into an approximate mechanical equilibrium under the current radii. This “expansion–equilibration” cycle continues until the radii recover the target values, thereby yielding a uniform and minimally overlapping layout. The iteration is terminated based on the usage rate (UR) defined in Section 4.2; the system is considered to have converged stably when the change in UR between successive iterations falls below a given threshold.
 
 \begin{definition}[Expansion System State]
 	For a given radius \(r\), the set of all circles' positions, velocities, and total external forces is defined as the expansion system state, denoted by
 	\[
 	S(r) = \left\{ \boldsymbol{p}_i(r), \boldsymbol{v}_i(r), \boldsymbol{F}_{\text{total},i}(r) \right\}_{i=1}^{n},
 	\]
 	where \(\boldsymbol{p}_i(r)\) represents the position of the \(i\)-th circle, \(\boldsymbol{v}_i(r)\) its velocity, and \(\boldsymbol{F}_{\text{total},i}(r)\) the total external force acting on it.
 \end{definition}
 
 The iterative process of the expansion system can be formulated as
 \begin{equation}
\left\{ \begin{array}{l}
	S\left( r_{k+1} \right) =\mathcal{O}\left( S\left( r_k \right) ,r_{k+1} \right)\\
	r_{k+1}=r_k+\Delta r_k\\
\end{array} \right. 
 \end{equation}
 where \(\mathcal{O}\) represents the optimization operator that transforms the system from the state at radius \(r_k\) to the state at radius \(r_{k+1}\). We employ an adaptive step size to update the circle radii, using the usage rate as the criterion for radius adjustment, as follows:
 \begin{equation}
 	\Delta r_{k+1}=\begin{cases}
 		2\Delta r_k&		\text{if\,\,}U\left( r_k+\Delta r_k \right) >U_{\text{th}}\,\,\text{and\,\,}C\geq C_{\text{th}}\\
 		\frac{1}{2}\Delta r_k&		\text{if\,\,}U\left( r_k+\Delta r_k \right) \leq U_{\text{th}}\\
 		\Delta r_k&		\text{otherwise}\\
 	\end{cases}
 \end{equation}
 Here, \( U(r) \) denotes the usage rate of the circles at radius \( r \), and \( U_{\text{th}} \) is the corresponding usage threshold. The counter \( C \) records the number of consecutive successful radius expansions, while \( C_{\text{th}} \) defines the acceleration threshold for step‑size doubling. This design enables the algorithm to accelerate expansion when the distribution exhibits a high usage rate, and to reduce the step size for fine‑grained search when usage rate is low.

\begin{algorithm}[H]
	\caption{Virtual-Force and Radius Iteration Process}
	\label{alg:radius_inflation}
	\begin{algorithmic}[1]
		\Require Convex polygon $P$, circles' number $N$, target radius $r_{\text{target}}$
		\Ensure Optimized expansion system $\mathcal{S}$
		\State $r_{\text{current}} \gets 0.1 \times r_{\text{target}}$ \Comment{Initial small radius}
		\State $\Delta r_k \gets r_{\text{target}} \times \alpha_{\text{inflate}}$ \Comment{Initial inflation step size}
		\State $\mathcal{S} \gets \text{initialize\_hexagonal\_grid}(P, N, r_{\text{current}})$
		\State $C \gets 0$
		\While{$\Delta r_k > \epsilon_{\text{inflate}}$} \Comment{Main inflation loop}
		\State $r_{\text{test}} \gets r_{\text{current}} + \Delta r_k$ \Comment{Try larger radius}
		\State $\mathcal{S}_{\text{test}} \gets \mathcal{S}.\text{copy}()$
		\State $\mathcal{S}_{\text{test}}.\text{radius} \gets r_{\text{test}}$
		\State $\mathcal{S}_{\text{test}} \gets \text{optimize\_positions}(\mathcal{S}_{\text{test}})$ \Comment{Optimize under new radius}
		\If{$\mathcal{S}_{\text{test}}.\text{usage\_rate} > U_\text{th}$} \Comment{Evaluate system state}
		\State $\mathcal{S} \gets \mathcal{S}_{\text{test}}$ \Comment{Accept new radius}
		\State $r_{\text{current}} \gets r_{\text{test}}$
		\State $C \gets C + 1$
		\If{$C \geq C_{\text{th}}$}
		\State $\Delta r_k \gets \Delta r_k \times 2$ \Comment{Accelerate inflation}
		\State $C \gets 0$
		\EndIf
		\Else
		\State $\Delta r_k \gets \Delta r_k / 2$ \Comment{Shrink step size}
		\State $C \gets 0$
		\EndIf
		\EndWhile
		\State \Return $\mathcal{S}$
	\end{algorithmic}
\end{algorithm}
 
 \subsection{Boundary Encirclement Strategy}
 Based on the structure-preserving initialization strategy and the constraints of virtual forces, the majority of  circles are concentrated within the interior region of the convex polygon. However, some circles inevitably overflow beyond the polygon boundary, necessitating the development of a mechanism to handle boundary overflow. To address this, we propose a boundary encircling strategy. The core idea is to relocate the overflowing circles along the boundary gradient direction to the nearest feasible positions.

We reformulate the original problem as minimizing the negative coverage ratio to make it compatible with gradient descent, as follows:
 \begin{equation}
 	J_{\text{cover}}(\boldsymbol{C}) = -\frac{A_{\text{cover}}(\boldsymbol{C})}{S_{P}}.
 \end{equation}
 
 \begin{definition}[Boundary Distance Term]
 	To discourage circles from exceeding the boundary, we define the boundary distance term
 	\[
 	J_{\text{boundary}}(\boldsymbol{C}) = \frac{1}{n} \sum_{i=1}^n \left( \min_{j} d_j(\boldsymbol{c}_i) - \alpha r \right)^2.
 	\]
 	where \(\alpha \in [1,\infty )\) controls the desired distance between a circle and the boundary, and \(d_j(\boldsymbol{c}_i)\) denotes the signed distance from the circle center \(\boldsymbol{c}_i\) to the \(j\)-th edge of the polygon. The condition \(\min_j d_j(\boldsymbol{c}_i) = \alpha r\) indicates that the center is positioned exactly \(\alpha r\) away from its closest polygon edge.
 \end{definition}
 
 To retrieve overflow circles and implement boundary-encircling optimization, we sequentially introduce two constraints: a boundary constraint and an inter-center distance constraint. For all $i,j$, the following inequality constraints must be satisfied:
 \begin{equation}
 	\left\{ \begin{array}{l}
 		g_j\left( \boldsymbol{c}_i \right) =-\left[ d_j\left( \boldsymbol{c}_i \right) -r \right] \le 0\\
 		h_{ij}\left( \boldsymbol{C} \right) =2r-||\boldsymbol{c}_i-\boldsymbol{c}_j||\le 0\\
 	\end{array} \right. .
 \end{equation}
The augmented Lagrangian function is then formulated as follows:
 \begin{equation}
 	\begin{split}
 		\mathcal{L}(\boldsymbol{C}, \lambda, \mu, \rho) 
 		&= J_{\text{cover}}(\boldsymbol{C}) + \beta J_{\text{boundary}}(\boldsymbol{C}) \\
 		&\quad + \sum_{i,j} \Big[ \mu_{ij} \max\big(0, g_j(\boldsymbol{c}_i)\big) 
 		+ \frac{\rho_b}{2} \max\big(0, g_j(\boldsymbol{c}_i)\big)^2 \Big] \\
 		&\quad + \sum_{i<j} \Big[ \lambda_{ij} \max\big(0, h_{ij}(\boldsymbol{C})\big) 
 		+ \frac{\rho_c}{2} \max\big(0, h_{ij}(\boldsymbol{C})\big)^2 \Big].
 	\end{split}
 \end{equation}
 
 \subsubsection{Normal Gradient}
 Differentiating the boundary constraint leads to the normal gradient:
 \begin{equation}
 	\nabla _{\bot}\mathcal{L} = -\nabla_{\boldsymbol{c}_i} \left[ \frac{\rho_b}{2} \max\big(0, g_j(\boldsymbol{c}_i)\big)^2 \right] 
 	= \begin{cases}
 		\rho _b\bigl( r-d_j\left( \boldsymbol{c}_i \right) \bigr) \cdot \frac{\boldsymbol{n}_j}{\|\boldsymbol{n}_j\|}, & \text{if } d_j\left( \boldsymbol{c}_i \right) < r\\
 		0, & \text{otherwise}
 	\end{cases}.
 \end{equation}
 
 This gradient provides an inward corrective force when the circle overflows($d_j(\boldsymbol{c}_i) < r$), and vanishes otherwise.
 
 \subsubsection{Tangential Gradient}
 
 To drive the overflow circles along the boundary for optimal placement, we incorporate a tangential gradient that promotes boundary encircling. Let \( k \) be the index of the edge closest to the circle center \(\boldsymbol{c}_i\), with unit normal vector \(\boldsymbol{n} = \boldsymbol{n}_k / \|\boldsymbol{n}_k\|\). The tangent vector \(\boldsymbol{t}\) satisfies \(\boldsymbol{t} \cdot \boldsymbol{n} = 0\).
 
 Differentiating the boundary distance term yields its gradient:
 \begin{equation}
 	\nabla_{\boldsymbol{c}_i} J_{\text{boundary}} = \frac{2}{n} \bigl( d_k(\boldsymbol{c}_i) - \alpha r \bigr) \boldsymbol{n}.
 \end{equation}
 Since $\nabla_{\boldsymbol{c}_i} J_{\text{boundary}}$ is collinear with the normal vector $\boldsymbol{n}$, its projection onto the tangential space is zero:
 \begin{equation}
 	\nabla_{\parallel} J_{\text{boundary}} = \boldsymbol{P}_{\parallel} \nabla_{\boldsymbol{c}_i} J_{\text{boundary}} = \boldsymbol{0},
 \end{equation}
 where $\boldsymbol{P}_{\parallel} = \boldsymbol{I} - \boldsymbol{n}\boldsymbol{n}^\top$ is tangential projection operator.
 
 Therefore, to generate a driving force along the boundary, we further introduce the concept of \textit{boundary encircling points}.
 
 \begin{definition}[Boundary Encircling Points]
 	Uniformly spaced boundary encircling points are generated along each polygon edge:
 	\begin{equation}
\boldsymbol{b}_{j,l}=\boldsymbol{v}_j+\frac{l-0.5}{n_j}\left( \boldsymbol{v}_{j+1}-\boldsymbol{v}_j \right) +\alpha r\cdot \boldsymbol{n}_j\left( l=1,\cdots ,n_j \right) ,
 	\label{eq:27}
 	\end{equation}
 	where the number of points on the \(j\)-th edge is defined as  
 	\[
 	n_j = \max\left(1, \Bigl\lfloor n \cdot \frac{\|\boldsymbol{v}_{j+1} - \boldsymbol{v}_j\|}{L}\Bigr\rfloor\right).
 	\]
 	Here, \(n\) is the total number of circles, and \(L\) denotes the perimeter of the polygon. The symbols are defined in Table ~\ref{tab:1}.
 \end{definition}
 
 \begin{table}[!ht]
 	\centering
 	\caption{Description of Boundary Encircling Point Parameters}
 	\begin{tabular}{cl}
 		\toprule[1.5pt]
 		\textbf{Symbol} & \textbf{Meaning} \\
 		\midrule[1pt]
 		$\boldsymbol{b}_{j,l}$ & Coordinates of the $l$-th boundary encircling point on the $j$-th edge. \\
 		$\boldsymbol{v}_j$, $\boldsymbol{v}_{j+1}$ & Coordinates of the $j$-th and $(j+1)$-th vertices of the polygon. \\
 		$n_j$ & Number of boundary encircling points allocated to the $j$-th edge. \\
 		$\frac{l-0.5}{n_j}$ & Normalized positional parameter with values in $(0,1)$. \\
 		$\alpha$ & Boundary offset coefficient ($\alpha \in [0,1]$)  \\
 		$r$ & Radius of the circles. \\
 		$\boldsymbol{n}_j$ & Unit inward normal vector of the $j$-th edge (pointing inside the polygon). \\
 		$\mathcal{B}$ & Set of all boundary encircling points: $\mathcal{B} = \{\boldsymbol{b}_{j,l} \mid \forall j,l\}$, \\
 		\bottomrule[1.5pt]
 	\end{tabular}
 	\label{tab:1}
 \end{table}
 
 $\frac{l-0.5}{n_j}$ ensures points are located strictly inside the edge. $\alpha$ controlls the degree of inward offset along the normal direction. $\mathcal{B}$ serves as candidate target positions for overflow circles. For each circle $C_i$, we should find its nearest boundary encircling point:
 \begin{equation}
 	\boldsymbol{b}_i^* = \arg\min_{\boldsymbol{b} \in \mathcal{B}} \|\boldsymbol{c}_i - \boldsymbol{b}\|.
 \end{equation}
 The tangential attractive force generated by boundary encircling as follows:
 \begin{equation}
 	\boldsymbol{F}_{\text{border}}(\boldsymbol{c}_i) = \gamma (\boldsymbol{b}_i^* - \boldsymbol{c}_i),
 \end{equation}
 where $\gamma > 0$ denotes the attractive coefficient. Then, We obtain the tangential gradient as the projection of this attractive force onto the tangential space:
 \begin{equation}
 	\nabla_{\parallel} \mathcal{L}_{\text{border}} = \boldsymbol{P}_{\parallel} \boldsymbol{F}_{\text{border}}(\boldsymbol{c}_i) 
 	= \gamma \boldsymbol{P}_{\parallel} (\boldsymbol{b}_i^* - \boldsymbol{c}_i).
 \end{equation}
 
 \subsubsection{Repulsive Gradient between Circles}
The repulsive gradient induced by inter-circle constraints is given by:
 
 \begin{equation}
 	\begin{split}
 		\nabla_{\text{circle}}\mathcal{L} &= 
 		\nabla_{\boldsymbol{c}_i}\left[ \frac{\rho_c}{2} \max\left(0, h_{ij}(\boldsymbol{C})\right)^2 \right] \\
 		&= \begin{cases}
 			\rho_c \bigl( 2r - \|\boldsymbol{c}_i - \boldsymbol{c}_j\| \bigr) \cdot 
 			\dfrac{\boldsymbol{c}_i - \boldsymbol{c}_j}{\|\boldsymbol{c}_i - \boldsymbol{c}_j\|} & 
 			\text{if } \|\boldsymbol{c}_i - \boldsymbol{c}_j\| < 2r, \\
 			0 & \text{otherwise}.
 		\end{cases}
 	\end{split}
 \end{equation}
 
 By integrating the above gradients, the update formula for circle center optimization when overflow occurs is given by:
 \begin{equation}
 	\boldsymbol{c}_i^{(t+1)} = \boldsymbol{c}_i^{(t)} - \eta \left[ \nabla_{\bot}\mathcal{L} + \nabla_{\parallel} \mathcal{L}_{\text{border}} + \sum_{j \neq i} \nabla_{\text{circle}}\mathcal{L} \right],
 \end{equation}
 where $\eta > 0$ denotes learning rate. The algorithm's convergence criterion is set as
 \begin{equation}
 	\max_i \|\boldsymbol{c}_i^{(t+1)} - \boldsymbol{c}_i^{(t)}\| < \epsilon,
 \end{equation}
where $\epsilon > 0$ is a predefined convergence threshold. For example, The boundary encircling strategy is illustrated in Fig. ~\ref{fig:8}.

\section{Conclusion and Future Work}

This paper proposes an Improved Quasi-Physical Algorithm for solving the optimal circular coverage in arbitrary convex polygons. The improvements are mainly summarized as follows:

(1) A structure-preserving initialization strategy based on regular hexagonal tiling via scaling and affine transformations is designed, which effectively enhances the quality of the initial solution and reduces the time required for subsequent iterative optimization;  

(2) A virtual force field incorporating friction and a radius-expansion optimization iteration model are constructed, expanding the global search space, reducing the degree of overlap among circles, and improving the quality of the final solution ultimately;  

(3) A boundary-surrounding strategy based on normal and tangential gradients is proposed. This strategy is used to retrieve circles that overflow the polygon boundary, allowing originally overflowing circles to re-enter the interior of the polygon for effective coverage, thereby significantly increasing the usage rate of circles and the coverage rate.  

Subsequently, this paper conducts small-scale and large-scale experiments on rectangles, regular convex polygons, and arbitrary irregular convex polygons. The experimental results show that the proposed algorithm significantly outperforms the latest existing meta-heuristic algorithms (ICGWO, ICM-MS, MIFSA, VGSOK) in the two key metrics of coverage rate and circle utilization rate under all test conditions, while also achieving the lowest performance in boundary adaptability. Although low boundary adaptability contributes less to coverage near the boundary, it helps prevent circles from overflowing the boundary, thereby improving both coverage rate and usage rate. In large-scale tests across all regions, the IQPD algorithm outperforms the other four algorithms in all metrics except for computation time and minimum gap, which are at moderate levels. In small-scale tests, the IQPD algorithm excels in the minimum gap metric, while its distribution quality and uniformity index are moderate. In all small-scale tests, the runtime of the IQPD algorithm is almost the lowest, and in all large-scale tests, its runtime remains moderate, with computation time never increasing significantly with polygon complexity. Therefore, in large-scale tests, the comprehensive advantages of the IQPD algorithm across multiple metrics are more pronounced.  

Although the proposed IQPD algorithm surpasses existing methods in several metrics, due to page limitations, we have not conducted an in-depth analysis of which improvement module contributes more to the final results, nor have we performed a more detailed theoretical analysis and ablation experiments on the algorithmic complexity, convergence, and parameter sensitivity of each module. This will be the first step in further refining this work in the future. In addition, the research in this paper is limited to convex polygons, whose convexity simplifies the problem modeling. However, real-world scenarios are often more complex, possibly involving non-convex regions with holes or concave corners. Therefore, constructing a general circle coverage mechanism applicable to arbitrary two-dimensional planar regions is an important direction for future work. If these issues are successfully addressed, further research could be conducted on coverage optimization problems with dynamic adjustment of circle radii and numbers, optimization of unequal circle coverage, and the densest packing problem in three-dimensional space. It is anticipated that the IQPD algorithm can be applied to practical problems like wireless network coverage. Further research in this direction would contribute valuable solutions to operational optimization and resource allocation.

 \section{Acknowledgements}
 
 The work was supported by National Natural Science Foundation of China(Grant No. 12371016, 11871083) and National Key R \& D Program of
 China(Grant No. 2020YFE0204200).

\bibliographystyle{elsarticle-num-names} 
 \bibliography{references}  
\end{document}